\documentstyle[12pt]{article}

\title{\bf Scale transformation, modified gravity, and Brans-Dicke theory}
\author{F. Darabi\thanks{e-mail:
f.darabi@azaruniv.edu}\\{\small Department of Physics, Azarbaijan
University of Tarbiat Moallem, 53714-161, Tabriz, Iran .} }
\begin{document}
\maketitle
\begin{abstract}
A model of Einstein-Hilbert action subject to the scale
transformation is studied. By introducing a dilaton field as a means
of scale transformation a new action is obtained whose Einstein
field equations are consistent with traceless matter with
non-vanishing modified terms together with dynamical cosmological
and gravitational coupling terms. The obtained modified Einstein
equations are neither those in $f(R)$ metric formalism nor the ones
in $f({\cal R})$ Palatini formalism, whereas the modified source
terms are {\it formally} equivalent to those of $f({\cal
R})=\frac{1}{2}{\cal R}^2$ gravity in Palatini formalism. The
correspondence between the present model, the modified gravity
theory, and Brans-Dicke theory with $\omega=-\frac{3}{2}$ is
explicitly shown, provided the dilaton field is condensated to its
vacuum state.

PACS: Keywords: Scale transformation; modified gravity; Brans-Dicke
theory.
\end{abstract}
\newpage
\section{Introduction}

Conformal invariance has played a key role in the study of $G$ and
$\Lambda$ varying theories raised by Dirac \cite{Dirac}. Bekenstein
was the first who introduced this possibility and tried to resolve
$G$-varying problem \cite{Bekenstein}. The main point in  this idea
is the assumption that the time variation of couplings is generated
by the dynamics of a cosmological scalar (dilaton) field. For
example, Damour {\it et al} have constructed a generalized
Jordan-Brans-Dicke model in which the dilaton field couples with
different strengths to the visible and dark matter \cite{Damour}. On
the other hand, Bertolami \cite{Bertolami} has introduced a model in
which both gravitational coupling and cosmological term are time
dependent. Conformal invariance implies that the gravitational
theory is invariant under local changes of units of length and time.
These local transformations relate different unit systems or
conformal frames via space time dependent conformal factors, and
these unit systems are dynamically distinct. This usually leads to
the variability of the fundamental constants. Recently, it is shown
that one can use this dynamical distinction between two unit systems
usually used in cosmology and particle physics to alleviate the
cosmological constant and coincidence problems \cite{Bisabr}.

Unlike the above models based on conformal invariance, the purpose
of present paper is to study a gravitational model in which {\it
scale transformations} play the key role in obtaining dynamical $G$
and $\Lambda$. A scale transformation is different from a conformal
transformation. A conformal transformation is viewed as
``stretching" all lengths by a space time dependent conformal
factor, namely a ``unit" transformation. But, a scale transformation
is just a rescaling of metric by a space time dependent conformal
factor, and all lengths are assumed to remain unchanged. This kind
of transformation is not a ``unit" transformation; it is just a
dynamical rescaling ( enlargement or contraction ) of a system. We
will take a non-scale invariant gravitational action with a {\it
cosmological constant} in which gravity couples minimally to a
dimensionless dilaton field, and matter couples to a metric which is
conformally related, through the dilaton field, to the gravitational
metric. Then by a scale transformation, through the dilaton field,
we obtain a new action in which gravity couples non-minimally to the
dilaton field and matter couples to the gravitational metric. The
Einstein equations reveal a cosmological term and a gravitational
coupling which are dynamically dependent on the dilaton field ( or
conformal factor ) satisfying the field equation with a Higgs type
potential. The symmetry breaking in this potential may occur for
positive cosmological constant and leads to the vacuum condensation
of the dilaton field. By putting this vacuum state of the dilaton
field in the Einstein-dilaton equations we obtain the modified
Einstein equations in the spirit of current $f(R)$ theories of
gravity \cite{Sotiriou}. Then, we study the characteristics of these
equations and examine the correspondence with equations obtained in
metric and Palatini formalisms in one hand, and Brans-Dicke theory
on the other hand. It is very appealing to study the cosmology of
this model and examine the correspondence with those recent
interesting works on the cosmology of modified gravity theory and
Brans-Dicke theory, for example the agegraphic dark energy density
in the $f(R)$ gravity and holographic dark energy density in the
Brans-Dicke theory \cite{Setare}.

\section{Einstein-Hilbert action and scale transformation}

We start with the following action\footnote{We will use the sign
convention $g_{\mu \nu}=diag(-, +, +, +)$ and units where $c=1$.}
\begin{equation}
S=\frac{1}{2\bar{\kappa}^2}\int  \sqrt{-g}[{R}-2\bar{\Lambda}+\alpha
g^{\mu \nu}\nabla_{\mu}\sigma \nabla_{\nu}\sigma]d^4x +
S_m(e^{2\sigma}{g}_{\mu \nu}),\label{1}
\end{equation}
where Einstein-Hilbert action with metric $g_{\mu \nu}$ is minimally
coupled to a dimensionless dilaton field $\sigma$, and the matter is
coupled to gravity with the metric $e^{2\sigma}g_{\mu \nu}$ which is
conformally related to the metric $g_{\mu \nu}$\footnote{The idea of
coupling matter to conformally related metrics has already been
proposed by some authors \cite{Damour}.}. The parameters
$\bar{\kappa}^2$ and $\bar{\Lambda}$ are gravitational coupling and
cosmological constants, respectively, and $\alpha$ is to be
determined later. Variation with respect to $g_{\mu\nu}$ and
$\sigma$ yields
\begin{equation}
G_{\mu \nu}+ \bar{\Lambda} g_{\mu \nu}=\bar{\kappa}^2\tilde{T}_{\mu
\nu} + \tau_{\mu \nu},\label{3}
\end{equation}
\begin{equation}
\Box \sigma=\frac{\bar{\kappa}^2}{\alpha}\tilde{T}, \label{5}
\end{equation}
where
\begin{equation}
\tilde{T}_{\mu \nu}=\frac{2}{\sqrt{-g}}\frac{\delta
S_m(e^{2\sigma}{g}_{\mu \nu})}{\delta g^{\mu \nu}},\label{6}
\end{equation}
\begin{equation}
\tau_{\mu \nu}=\alpha(\frac{1}{2}g_{\mu \nu}\nabla_{\gamma}\sigma
\nabla^{\gamma}\sigma-\nabla_{\mu}\sigma \nabla_{\nu}\sigma
),\label{7}
\end{equation}
and $\tilde{T}$ is the $g_{\mu \nu}$ trace of the energy-momentum
tensor $\tilde{T}_{\mu \nu}$. Now, we introduce the scale
transformations
\begin{equation}
g_{\mu \nu} \rightarrow \Omega^2g_{\mu \nu}, \label{77}
\end{equation}
\begin{equation}
\sqrt{-g}\rightarrow \Omega^4 \sqrt{-g}, \label{8}
\end{equation}
\begin{equation}
{R}\rightarrow
\Omega^{-2}{R}+6\Omega^{-3}\nabla_{\mu}\nabla_{\nu}\Omega g^{\mu
\nu}.\label{9}
\end{equation}
where $\Omega=e^{-\sigma}$. The action (\ref{1}) then becomes
\begin{equation}
 S=\frac{1}{2\bar{\kappa}^2}\int \sqrt{-g}[{R}\Omega^2+6\Omega\Box \Omega
 -2\bar{\Lambda}\Omega^4 +\alpha g^{\mu \nu}\nabla_{\mu}\Omega
 \nabla_{\nu}\Omega]d^4x
 + S_m(g_{\mu \nu}),\label{10}
\end{equation}
where gravity couples non-minimally to the dilaton field and matter
couples to the gravitational metric $g_{\mu \nu}$. The field
equations are obtained by variation of (\ref{10}) with respect to
the fields $g_{\mu \nu}$ and $\Omega$ as
\begin{equation}
G_{\mu \nu}+ \Omega^2\bar{\Lambda} g_{\mu
\nu}=\Omega^{-2}\bar{\kappa}^2 {T}_{\mu \nu} + \tau_{\mu
\nu}(\Omega),\label{11}
\end{equation}
and
\begin{equation}
\Box
\Omega-\frac{1}{(\alpha-6)}({R}-4\bar{\Lambda}\Omega^2)\Omega=0,
\label{12}
\end{equation}
where
\begin{equation}
\tau_{\mu \nu}(\Omega)=(\alpha-6)\Omega^{-2} [\frac{1}{2}g_{\mu
\nu}\nabla_{\lambda}\Omega \nabla^{\lambda}\Omega-\nabla_{\mu}\Omega
\nabla_{\nu}\Omega]-\Omega^{-2}[g_{\mu \nu}\Box -
\nabla_{\mu}\nabla_{\nu}]\Omega^2.\label{13}
\end{equation}
Manipulating the last term in (\ref{13}) and taking the $g_{\mu
\nu}$ trace of Eq.(\ref{11}) gives
\begin{equation}
-{R}+4\Omega^2\bar{\Lambda}=\Omega^{-2}\bar{\kappa}^2
{T}+(\alpha-6)\Omega^{-2}\nabla_{\lambda}\Omega
\nabla^{\lambda}\Omega -6\Omega^{-2}(\nabla_{\lambda}\Omega
\nabla^{\lambda}\Omega+\Omega \Box\Omega).\label{133}
\end{equation}
It is easily seen that Eqs.(\ref{12}) and (\ref{133}) are just
consistent for $\alpha=12$ which is a free parameter in the action.
This value for $\alpha$ then leads the two equations to take the
following forms
\begin{equation}
\Box \Omega-\frac{1}{6}({R}-4\bar{\Lambda}\Omega^2)\Omega=0,
\label{122}
\end{equation}
and
\begin{equation}
\Box
\Omega-\frac{1}{6}({R}-4\bar{\Lambda}\Omega^2)\Omega=\Omega^{-2}\bar{\kappa}^2
{T}, \label{1333}
\end{equation}
respectively. It turns out that Eqs.(\ref{12}) and (\ref{133}) are
consistent just for traceless energy-momentum tensors, namely $T=0$.
One may rewrite equation (\ref{11}) as
\begin{equation}
G_{\mu \nu}+ \Lambda g_{\mu \nu}=\kappa^2 {T}_{\mu \nu} + \tau_{\mu
\nu}(\Omega),\label{14}
\end{equation}
where $\Lambda=\Omega^2\bar{\Lambda}$ and
$\kappa^2=\Omega^{-2}\bar{\kappa}^2$. If we compare Eq.(\ref{12}),
for $\alpha=12$, with the general form of the $\Omega$ field
equation, namely $\Box\Omega+\frac{dV}{d\Omega}=0$, we may infer the
Higgs type potential for $\Omega$ as
\begin{equation}
V(\Omega)=-\frac{1}{12}({R}-2\bar{\Lambda}\Omega^2)\Omega^2.\label{155}
\end{equation}
For a given positive Ricci scalar, a negative $\bar{\Lambda}$ leads
to vanishing minimum for the conformal factor, namely
$\Omega_{min}=0$. This case is a failure of conformal transformation
with zero cosmological constant $\Omega^2\bar{\Lambda}$, so is not
physically viable. The nonvanishing minimum of this potential is
obtained for a positive $\bar{\Lambda}$ as
\begin{equation}
\Omega^2_{min}=\frac{R}{4\bar{\Lambda}}>0. \label{15}
\end{equation}
Putting this value of $\Omega_{min}$ in Eq.(\ref{14}) leads to the
generalized Einstein field equation
\begin{equation}
G_{\mu \nu}= -\frac{{R}}{4} g_{\mu \nu}+\frac{4\bar{\Lambda}
\bar{\kappa}^2}{R} {T}_{\mu \nu}+\frac{6}{{R}}[\frac{1}{2}g_{\mu
\nu}(\nabla_{\lambda}\sqrt{{R}})
(\nabla^{\lambda}\sqrt{{R}})-(\nabla_{\mu}\sqrt{{R}})
(\nabla_{\nu}\sqrt{{R}})]-\frac{1}{{R}}[g_{\mu \nu}\Box -
\nabla_{\mu}\nabla_{\nu}]{R}.\label{16}
\end{equation}
The important results are as follows:\\
1) When $\Omega$ or ${R}$ is constant, the theory reduces to the
standard GR with modified cosmological and gravitational constants,
$\Lambda=\Omega^2\bar{\Lambda}$,
$\kappa^2=\Omega^{-2}\bar{\kappa}^2$, respectively.\\
2) The field equations are consistent just for the matter fields for
which $T = 0$. The model then reduces to GR with dynamical
cosmological term $\frac{{R}}{4} g_{\mu \nu}$ and coupling term
$\frac{4\bar{\Lambda} \bar{\kappa}^2}{R}$, and extra terms
containing derivatives of the Ricci scalar.

\section{Correspondence with modified gravity}

The Einstein equations can be derived using the Palatini formalism,
i.e., an independent variation with respect to the metric and an
independent connection. The Riemann tensor and the Ricci tensor are
also constructed with the independent connection and the metric is
not needed to obtain the latter from the former. So, in order to
make a difference with metric formalism, we shall use ${\cal R}_{\mu
\nu}$ and ${\cal R}$ instead of $R_{\mu \nu}$ and $R$, respectively.
In this section we briefly review the $f({\cal R})$ gravity in
Palatini formalism \cite{Sotiriou}. The action in the Palatini
formalism takes the form
\begin{equation}
S=\frac{1}{2\bar{\kappa}^2}\int  \sqrt{-g}f({\cal R})d^4x+
S_m({g}_{\mu \nu}).\label{17}
\end{equation}
Note that the matter action is assumed to depend only on the metric
and the matter fields and not on the independent connection. This
assumption is crucial for the derivation of Einstein's equations
from the action (\ref{17}) and is the main feature of the Palatini
formalism. Varying the action (\ref{17}) independently with respect
to the metric and the connection, respectively, and using the
formula
\begin{equation}
\delta{\cal R}_{\mu \nu}=\bar{\nabla}_{\lambda}\delta
\Gamma^{\lambda}_{\mu \nu}-\bar{\nabla}_{\nu}\delta
\Gamma^{\lambda}_{\mu \lambda},\label{18}
\end{equation}
yields
\begin{equation}
f^{\prime}({\cal R}){\cal R}_{(\mu \nu)}-\frac{1}{2}f({\cal
R})g_{\mu \nu}=\bar{\kappa}^2 {T}_{\mu \nu},\label{19}
\end{equation}
\begin{equation}
-\bar{\nabla}_{\lambda}(\sqrt{-g}f^{\prime}({\cal R})g^{\mu
\nu})+\bar{\nabla}_{\sigma}[\sqrt{-g}f^{\prime}({\cal R})g^{\sigma
(\mu }]\delta_{\lambda}^{\nu)}=0,\label{20}
\end{equation}
where $\bar{\nabla}$ denotes the covariant derivative defined with
the independent connection $\Gamma^{\lambda}_{\mu \nu}$ and $(\mu
\nu)$ denotes symmetrization over the indices $\mu, \nu$. Taking the
trace of Eq.(\ref{20}) gives rise to
\begin{equation}
\bar{\nabla}_{\sigma}[\sqrt{-g}f^{\prime}({\cal R})g^{\sigma \mu
}]=0,\label{21}
\end{equation}
by which the field equation (\ref{20}) becomes
\begin{equation}
\bar{\nabla}_{\lambda}(\sqrt{-g}f^{\prime}({\cal R})g^{\mu
\nu})=0.\label{22}
\end{equation}
One may obtain some useful manipulations of the field equations.
Taking the trace of Eq.(\ref{19}) yields an algebraic equation in
${\cal R}$
\begin{equation}
f^{\prime}({\cal R}){\cal R}-{2}f({\cal R})=\bar{\kappa}^2
{T}.\label{23}
\end{equation}
For traceless energy-momentum tensors $T = 0$, the Ricci scalar
${\cal R}$ will therefore be a constant root of the equation
\begin{equation}
f^{\prime}({\cal R}){\cal R}-{2}f({\cal R})=0.\label{24}
\end{equation}
It is apparent from Eq.(\ref{23}) that if $f({\cal R})\propto {\cal
R}^2$ then only conformally invariant matter, for which $T = 0$ is
identically satisfied, can be coupled to gravity \cite{Ferraris}.
One may define a metric conformal to $g_{\mu \nu}$ as
\begin{equation}
h_{\mu \nu}=f^\prime({\cal R})g_{\mu \nu},\label{25}
\end{equation}
for which it is easily obtained that
\begin{equation}
\sqrt{-h}h^{\mu \nu}=\sqrt{-g}f^\prime({\cal R})g^{\mu
\nu}.\label{26}
\end{equation}
Equation (\ref{22}) is then the compatibility condition of the
metric $h_{\mu \nu}$ with the connection $\Gamma^{\lambda}_{\mu
\nu}$ and can be solved algebraically to give
\begin{equation}
\Gamma^{\lambda}_{\mu \nu}=h^{\lambda \sigma}(\partial_{\mu} h_{\nu
\sigma}+\partial_{\nu} h_{\mu \sigma}-\partial_{\sigma} h_{\mu
\nu}).\label{27}
\end{equation}
Under conformal transformations (\ref{25}), the Ricci tensor and its
contracted form with $g^{\mu \nu}$ transform, respectively, as
\begin{equation}
{\cal R}_{\mu \nu}=R_{\mu \nu}+\frac{3}{2}\frac{1}{(f^\prime({\cal
R}))^2}(\nabla_{\mu}f^\prime({\cal R}))(\nabla_{\nu}f^\prime({\cal
R}))-\frac{1}{(f^\prime({\cal
R}))}(\nabla_{\mu}\nabla_{\nu}-\frac{1}{2}g_{\mu
\nu}\Box)f^\prime({\cal R}),\label{28}
\end{equation}
\begin{equation}
{\cal R}=R+\frac{3}{2}\frac{1}{(f^\prime({\cal
R}))^2}(\nabla_{\mu}f^\prime({\cal R}))(\nabla^{\mu}f^\prime({\cal
R}))+\frac{3}{(f^\prime({\cal R}))}\Box f^\prime({\cal
R}).\label{29}
\end{equation}
Note the difference between ${\cal R}$ and the Ricci scalar of
$h_{\mu \nu}$ due to the fact that $g^{\mu \nu}$ is used here for
the contraction of ${\cal R}_{\mu \nu}$. Substituting
Eqs.(\ref{28}), (\ref{29}) in the field equation (\ref{19}), one
obtains
$$
G_{\mu \nu}=\frac{\bar{\kappa}^2}{f^\prime({\cal R})}T_{\mu
\nu}-\frac{1}{2}g_{\mu \nu}\left({\cal R}-\frac{f({\cal
R})}{f^\prime({\cal R})}\right)+\frac{1}{f^\prime({\cal
R})}(\nabla_{\mu}\nabla_{\nu}-g_{\mu \nu}\Box)f^\prime({\cal R})
$$
\begin{equation}
-\frac{3}{2}\frac{1}{(f^\prime({\cal
R}))^2}[(\nabla_{\mu}f^\prime({\cal R}))(\nabla_{\nu}f^\prime({\cal
R}))-\frac{1}{2}g_{\mu \nu}(\nabla_{\lambda}f^\prime({\cal
R}))(\nabla^{\lambda}f^\prime({\cal R}))].\label{30}
\end{equation}
In fact, since Eq.(\ref{23}) relates ${\cal R}$ algebraically with
$T$, and that we have an explicit expression for
$\Gamma^{\lambda}_{\mu \nu}$ in terms of ${\cal R}$ and $h_{\mu
\nu}$ (or $g_{\mu \nu}$) we can in principle eliminate the
independent connection from the field equations and express them
only in terms of the metric and the matter fields. Therefore, both
sides of Eq.(\ref{30}) depend only on the metric and the matter
fields and the theory has been reduced to the form of GR with a
modified source.\\
The results of this Palatini's method for $f({\cal R})$
gravity are as follows:\\
1) When $f({\cal R}) = {\cal R}$, the theory reduces to the
standard GR.\\
2) For the matter fields for which $T = 0$, the Ricci scalar ${\cal
R}$ and consequently $f({\cal R})$ and $f^\prime({\cal R})$ are
constants (due to Eq.(\ref{24})) and the theory reduces to GR with a
cosmological constant and a modified coupling constant
$\bar{\kappa}^2/f^\prime$. If ${\cal R}_0$ is the value of ${\cal
R}$ when $T = 0$, then the value of the cosmological constant is
\begin{equation}
\frac{1}{2}\left({\cal R}_0-\frac{f({\cal R}_0)}{f^\prime({\cal
R}_0)}\right)=\frac{{\cal R}_0}{4},\label{31}
\end{equation}
where use has been made of Eq.(\ref{24}).\\
3) In the general case $T \neq 0$, the modified source on the right
hand side of Eq.(\ref{30}) includes derivatives of the stress-energy
tensor which are implicit in the last two terms, since $f^\prime$ is
in practice a function of $T$, namely $f^\prime = f^\prime({\cal
R})$ and ${\cal R}={\cal R}(T )$.\\
It is easily shown that one may recover Eq.(\ref{16}) through an
equation similar to Eq.(\ref{30}) in which $f({\cal R})$ is replaced
by $f(R)$ as
$$
G_{\mu \nu}=\frac{\bar{\kappa}^2}{f^\prime(R)}T_{\mu
\nu}-\frac{1}{2}g_{\mu \nu}\left( R-\frac{f(R)}{f^\prime(
R)}\right)+\frac{1}{f^\prime(R)}(\nabla_{\mu}\nabla_{\nu}-g_{\mu
\nu}\Box)f^\prime(R)
$$
\begin{equation}
-\frac{3}{2}\frac{1}{(f^\prime(R))^2}[(\nabla_{\mu}f^\prime(
R))(\nabla_{\nu}f^\prime(R))-\frac{1}{2}g_{\mu
\nu}(\nabla_{\lambda}f^\prime(R))(\nabla^{\lambda}f^\prime(R))],\label{311}
\end{equation}
provided that
\begin{equation}
f(R)=\frac{1}{2}R^2.\label{32}
\end{equation}
Therefore, there is a {\it formal} correspondence between equations
(\ref{30}) and (\ref{311}). Note that equation (\ref{311}) is
neither the same as equation (\ref{30}) in Palatini formalism
(according to (\ref{29}), ${\cal R}$ and $R$ are not the same) nor
the same as equation which is obtained in metric formalism as
\cite{Sotiriou}
\begin{equation}
G_{\mu \nu}=\frac{\bar{\kappa}^2}{f^\prime(R)}T_{\mu
\nu}+\frac{1}{2}g_{\mu \nu}\frac{[f(R)-R
f^\prime(R)]}{f^\prime(R)}+\frac{1}{f^\prime(R)}(\nabla_{\mu}\nabla_{\nu}-g_{\mu
\nu}\Box)f^\prime(R).
\end{equation}

\section{Correspondence with Brans-Dicke theory}

The $f({\cal R})$ gravity action coupled with matter is given by
(\ref{17})
$$
S=\frac{1}{2\bar{\kappa}^2}\int  \sqrt{-g}f({\cal R})d^4x+
S_m({g}_{\mu \nu}).
$$
By introducing a new auxiliary field $\chi$, the dynamically
equivalent action is rewritten \cite{Sotiriou}
\begin{equation}
S=\frac{1}{2\bar{\kappa}^2}\int
\sqrt{-g}[f(\chi)+f^\prime(\chi)({\cal R}-\chi)]d^4x+ S_m({g}_{\mu
\nu}).\label{33}
\end{equation}
Variation with respect to $\chi$ yields the equation
\begin{equation}
f^{\prime \prime}(\chi)({\cal R}-\chi)=0.\label{34}
\end{equation}
Redefining the field $\chi$ by $\phi = f^\prime(\chi)$ and
introducing  $V(\phi)=\chi(\phi)\phi-f(\chi(\phi))$ the action takes
the form
\begin{equation}
S=\frac{1}{2\bar{\kappa}^2}\int \sqrt{-g}[\phi{\cal R}-V(\phi)]d^4x+
S_m({g}_{\mu \nu}).\label{35}
\end{equation}
Now, we may use Eq.(\ref{29}) which relates ${\cal R}$ and $R$. To
this end, we use $\phi = f^\prime(\chi)$ subject to the constraint
(\ref{34}) so that we may replace $f^\prime({\cal R})$ by $\phi$ in
Eq.(\ref{29}) to obtain
\begin{equation}
{\cal
R}=R+\frac{3}{2}\frac{1}{(\phi)^2}(\nabla_{\mu}\phi)(\nabla^{\mu}\phi)+\frac{3}{(\phi)}\Box
\phi.
\end{equation}
Therefore, the action (\ref{35}) modulo surface terms obtained by
$\frac{3}{(\phi)}\Box \phi$ can be rewritten as
\begin{equation}
S=\frac{1}{2\bar{\kappa}^2}\int \sqrt{-g}\left(\phi
R+\frac{3}{2\phi}\nabla_{\mu}\phi
\nabla^{\mu}\phi-V(\phi)\right)d^4x+ S_m({g}_{\mu \nu}).\label{36}
\end{equation}
This is the action of a Brans-Dicke theory with Brans-Dicke
parameter $\omega =-\frac{3}{2}$. The generalized Einstein equation
obtained by variation of the action (\ref{36}) with respect to
$g_{\mu \nu}$ is
$$
G_{\mu \nu}=\frac{\bar{\kappa}^2}{\phi}T_{\mu
\nu}-\frac{3}{2\phi^2}\left(\nabla_{\mu}\phi
\nabla_{\nu}\phi-\frac{1}{2}g_{\mu \nu}\nabla_{\lambda}\phi
\nabla^{\lambda}\phi\right)
$$
\begin{equation}
+\frac{1}{\phi}(\nabla_{\mu}\nabla_{\nu}-g_{\mu
\nu}\Box)\phi-\frac{V(\phi)}{2\phi}g_{\mu \nu}.\label{37}
\end{equation}
It is easily shown that Eq.(\ref{16}) corresponds to Eq.(\ref{37})
provided that
\begin{equation}
\phi \equiv \Omega^2_{min},\:\:\:\: V(\phi)\equiv
12V(\Omega^2_{min})=\frac{R^2}{8\bar{\Lambda}}.\label{38}
\end{equation}
\newpage
\section*{Conclusion}

In this paper, we studied a model of scale transformation imposed on
a gravitational action coupled with matter which is not scale
invariant due to the presence of dimensional gravitational and
cosmological constants. By choosing a dilaton field as a means of
scale transformation we obtained a new action whose field equations
are consistent for traceless matter and reveal dynamical
cosmological term and gravitational coupling. We then examined the
correspondence between the present model, the modified gravity
theory, and Brans-Dicke theory.

First, we showed that the modified source terms in the obtained
Einstein equation may be {\it formally} considered as equivalent to
those of $f({\cal R})=\frac{1}{2}{\cal R}^2$ gravity in Palatini
formalism, but with two differences: 1) Unlike the $f({\cal R})$
gravity for a traceless matter $T=0$ which reduces to GR with
cosmological and modified coupling {\it constants}, the present
modified Einstein equation subject to $T=0$ has the advantage of
non-vanishing modified terms together with {\it dynamical}
cosmological and gravitational coupling terms. In other words,
unlike the $f({\cal R})$ gravity, the traceless matter in the
present model does not reduce to GR and can still be considered as
general as those matters with $T\neq 0$ which are studied in
$f({\cal R})$ gravity, 2) Unlike the $f({\cal R})$ gravity in
Palatini formalism where the modified source terms depend on the
quantity ${\cal R}$ which is not the Ricci scalar of $h_{\mu \nu}$,
the modified source terms here depend on the Ricci scalar $R$
constructed by the metric $g_{\mu \nu}$. Note that the above
correspondence is obtained provided the $\Omega$ field is
condensated to its vacuum state $\Omega_{min}$. Overall, the
generalized Einstein equation for traceless matter obtained here
through scale transformation is neither an equation in $f(R)$ metric
formalism nor the one in $f({\cal R})$ Palatini formalism. Finally,
we showed the present model corresponds to a Brans-Dicke theory with
Brans-Dicke parameter $\omega =-\frac{3}{2}$.

\section*{Acknowledgment}

This work  has been supported by the Research office of Azarbaijan
University of Tarbiat Moallem, Tabriz, Iran.

\end{document}